# Arcsecond-resolution $^{12}$CO mapping of the yellow hypergiants IRC +10420 and AFGL 2343 ⋆

A. Castro-Carrizo[1], G. Quintana-Lacaci[2], V. Bujarrabal[2], R. Neri[1], and J. Alcolea[3]

[1] Institut de RadioAstronomie Millimétrique, 300 rue de la Piscine, 38406 Saint Martin d'Hères, France
   e-mail: `(ccarrizo,neri)@iram.fr`
[2] Observatorio Astronómico Nacional (IGN), Apdo. 112, E-28803 Alcalá de Henares, Spain
   e-mail: `(g.quintana,v.bujarrabal)@oan.es`
[3] Observatorio Astronómico Nacional (IGN), c/ Alfonso XII nº13, E-28014 Madrid, Spain
   e-mail: `j.alcolea@oan.es`

**ABSTRACT**

*Context.* IRC +10420 and AFGL 2343 are the unique, known yellow hypergiants (YHGs) presenting a heavy circumstellar envelope (CSE).
*Aims.* We aim to study the morphology, exceptional kinematics, and excitation conditions of their CSEs, and the implications for mass-loss processes.
*Methods.* We have mapped the $^{12}$CO $J$=2−1 and 1−0 emission in these YHGs with the IRAM Plateau de Bure interferometer and the 30m telescope. We developed LVG models in order to analyze their circumstellar characteristics.
*Results.* The maps show that the overall shape of both CSEs is approximately spherical, although they also reveal several aspherical features. The CSE around IRC +10420 shows a rounded extended halo surrounding a bright inner region, with both components presenting aspherical characteristics. It presents a brightness minimum at the center. The envelope around AFGL 2343 is a detached shell, showing spherical symmetry and clumpiness at a level of ∼ 15% of the maximum brightness. The envelopes expand isotropically at ∼ 35 km s$^{-1}$, about two or three times faster than typical CSEs around AGB stars. High temperatures (∼ 200 K) are derived for the innermost regions in IRC +10420, while denser and cooler (∼ 30 K) gas is found in AFGL 2343.
*Conclusions.* The mass-loss processes in these YHGs have been found to be similar. The deduced mass-loss rates (∼ 10$^{-4}$–10$^{-3}$ $M_\odot$ yr$^{-1}$) are much higher than those obtained in AGB stars, and they present significant variations on time scales of ∼ 1000 yr.

## 1. Introduction

Yellow hypergiants (YHGs) are among the most luminous and massive stars (see, as general references, de Jager 1998; Jones et al. 1993; and Humphreys 1991). They are thought to have luminosities in the range 5.3 ≤ log$L$ ($L_\odot$) ≤ 5.9 and initial masses higher than ∼ 20 $M_\odot$. These objects are post-red supergiants that are undergoing a poorly known, but likely complex evolution. Some theories propose that some YHGs may evolve redwards in the future; but the stellar temperature is rapidly increasing in at least a few of them. For instance, the spectral type of the hypergiant IRC +10420 has changed from F8Ia to A5Ia in just 20 yr (Klochkova et al. 1997).

It is thought that during the red and yellow phases, these heavy stars eject as much as one half of their initial mass (e.g. Maeder & Meynet 1988; de Jagger 1998). This mass loss should then be determinant in their subsequent evolution, eventually leading to a supernova explosion. However, most of the well studied YHGs only show faint traces of circumstellar material. Massive circumstellar envelopes have been detected in molecular line emission, dust-scattered light, and IR emission, only around IRC +10420 (= IRAS 19244+1115) and AFGL 2343 (= IRAS 19114+0002 = HD 179821); see Hawkins et al. (1995), Meixner et al. (1999), Bujarrabal et al. (1992, 2001), Neri et al. (1998), and Castro-Carrizo et al. (2001a). From $^{13}$CO mm-wave single-dish data, Bujarrabal et al. (2001) and Castro-Carrizo et al. (2001a) measured very high circumstellar masses, of a few $M_\odot$ (assuming distances compatible with the high mass and luminosity expected in YHGs). CO lines also reveal quite high expansion velocities, ∼ 35 km s$^{-1}$. Taking into account that the ejection time is expected to be < 10$^4$ yr, the mass-loss rate at which such molecular shells were formed must be extremely high, ∼ 10$^{-3}$ $M_\odot$ yr$^{-1}$. Other properties of these molecular shells, like their shape and chemical

---

*Send offprint requests to*: ccarrizo@iram.fr
⋆ Based on observations carried out with the IRAM Plateau de Bure Interferometer. IRAM is supported by INSU/CNRS (France), MPG (Germany), and IGN (Spain).

composition, have not been studied very closely. We do not know if such massive ejections occur in most YHGs, during discrete events (or at least in strongly variable processes), or whether they just appear in some objects. Therefore, we cannot assess yet, from an observational point of view, to what extent mass ejection plays a major role in the evolution of YHGs as a whole, and not only in a few objects, which could be following a rather different (and perhaps faster) evolution.

The structure of the envelopes around IRC +10420 and AFGL 2343 is not well known yet. Accurate maps of the IR continuum from AFGL 2343 (Hawkins et al. 1995; Jura & Werner 1999) show a clearly detached shell about 5″ in size. The detached shell is also conspicuous in images of polarized dust-scattered light (Gledhill et al. 2001a). Claussen (1993) and Gledhill et al. (2001b) mapped the OH maser emission in this source, finding a large number of spots distributed over a shell that is very similar to the one found in the IR.

The information obtained on the structure of the envelope around IRC +10420 is, however, controversial. From IR imaging and SED analysis, it has been concluded that a large amount of material exists at angular distances smaller than 1″ from the star (Lipman et al. 2000; Blöcker et al. 1999; Oudmaijer et al. 1996). A similar result was deduced by Teyssier et al. (2006) from the relatively intense CO emission in high-$J$ transitions. However, HST imaging reveals that the amount of mass in regions closer than ∼ 1-2″ −though not negligible− is smaller than expected, suggesting a decrease in the mass-loss rate in the last ∼ 1000 yr (Humphreys et al. 1997). This is consistent with observations in thermal SiO emission (Castro-Carrizo et al. 2001a) that show a shell of ∼ 5″ in size and ∼ 1″.5 in inner radius.

This paper presents a study of the envelopes around IRC +10420 and AFGL 2343. We have mapped their emission in the $^{12}$CO $J$=2−1 and 1−0 lines with spatial resolutions ∼ 1″. Models of molecular line emission have been developed to disentangle the mass ejection history of these objects. As we will see from our analysis with an LVG code, $^{12}$CO emission is well-suited to measuring the mass of the different circumstellar components and, therefore, the rates of the mass-loss events that gave rise to them.

Because of the warm photosphere, the presence of a circumstellar envelope and the bluewards evolution, IRC +10420 and AFGL 2343 have sometimes been associated with or classified as post-AGB stars; see discussions by e.g. Josselin & Lèbre (2001), Reddy & Hrivnak (1999), and Kastner & Weintraub (1995). In fact, even assuming very high initial masses, if the stellar temperature of an object like IRC +10420 continues increasing, the surrounding shell will soon be ionized by the stellar radiation and become a planetary nebula with an extremely high luminosity. This discussion probably persists because the distance, mainly for AFGL 2343, is poorly known. From Hipparcos parallax measurements, the distance to AFGL 2343 is ∼ 5.6 kpc; the poor quality of these data led Josselin & Lèbre (2001) to propose that this object could be a "normal" PPN at a shorter distance. The distance to IRC +10420 has been carefully studied (see de Jager 1998, Jones et al. 1993), however, and we assume it to be ∼ 5 kpc.

We will assume in this paper that IRC +10420 and AFGL 2343 are both hypergiants and therefore the mentioned distances apply. As we will see, our molecular data tend to confirm this hypothesis, since the properties deduced from the lines in these objects are quite similar among them and significantly different from those of PPNe and CSEs of AGB stars.

## 2. Observations and imaging of $^{12}$CO $J$=2−1 and 1−0

We have mapped the emission of the transitions $^{12}$CO $J$=1−0 and $J$=2−1 in the yellow hypergiants (YHGs) AFGL 2343 and IRC +10420. Observations were performed with the Plateau de Bure interferometer. Data were also obtained with the 30m telescope in order to recover the flux filtered out by the array.

### 2.1. Observations with the Plateau de Bure interferometer

We observed the emission of the rotational transitions of $^{12}$CO $J$=2−1 at 1.3 mm (230.538 GHz) and $J$=1−0 at 2.6 mm (115.2712 GHz) in IRC +10420 and AFGL 2343 with the IRAM interferometer at Plateau de Bure (PdB, France). The interferometer consists of 6 antennas of 15 m in diameter with dual-band SIS heterodyne receivers. Observations of both sources were carried out in the so-called track-sharing mode by observing the two sources in one track in configurations 6Cp (in March 2002) and 6Dp (in March 2003); projected baselines ranged from 16 m to 196 m. IRC +10420 was observed at the position coordinates (J2000) $19^h26^m48\overset{s}{.}10$, +11°21′17″.0, and AFGL 2343 at $19^h13^m58\overset{s}{.}60$, +00°07′32″.00. 3C 273 and MWC 349 were observed to calibrate the absolute flux, 1923+210, 1749+096, and 1741−038 to calibrate visibility phases and amplitudes. The accuracy of the flux calibration is within 10% at 3 mm and 20% at 1 mm. The calibration and data analysis were performed in the standard way using the GILDAS[1] software package.

No continuum emission was detected in AFGL 2343, neither at the frequency of CO $J$=1−0 nor at that of $J$=2−1 above a noise (rms) level of 0.6 and 0.8 mJy beam$^{-1}$, respectively. For IRC +10420, however, marginally resolved, continuum emission was detected at 2.6 mm at a level of 8.5 mJy beam$^{-1}$ (14 × rms, the integrated flux being 12 mJy) and at 1.3 mm at a level of 13 mJy beam$^{-1}$ (17 × rms, the integrated flux equal to 26 mJy). These values were obtained by averaging all the channels with no line emission, up to a total bandwidth of 0.5 GHz at 3mm and 1 GHz at 1 mm.

Finally, in order to verify the good quality of the calibration, it was checked that no elongation or halo exist around the calibrators mapped from the data of each track. Therefore, it is not expected to have any spurious contribution from the calibration above the dynamic range.

---

[1] See http://www.iram.fr/IRAMFR/GILDAS for more information about the GILDAS software.

## 2.2. Merging with short-spacing 30m-telescope data

We performed on-the-fly (OTF) observations of the $^{12}$CO $J$=2–1 and 1–0 emission in IRC +10420 with the 30m IRAM telescope located at Pico Veleta (PV, Spain). We aimed at recovering the short-spacing data filtered out by the interferometer, i.e. with $uv$-radii smaller than 15m. The observations were carried out in May 2005, on two consecutive days. In order to verify the calibration of our observations, line profiles at the center of IRC +10420 and of the CO-bright post-AGB star CRL 2688 were obtained on both days. It is worth noting that the 3mm and 1mm profiles were found to be, respectively, 6% and 9% stronger than those in Bujarrabal et al. (2001, hereafter B01). The OTF data were analyzed with the CLASS software package and merged to the PdB visibilities. A field of 80″ containing the whole circumstellar emission was mapped. In Fig. 1, CO $J$ =2–1 short-spacing and interferometric visibilities of IRC +10420 are plotted. By comparison, the flux obtained when pointing with the 30m telescope at the nebular center is shown, indicating the need to map a larger field to recover the flux from the outermost parts of the nebula. The fitting of the amplitudes of the short-spacing and the interferometric visibilities is satisfactory at the $uv$−plane interface (at $uv$−radius = 15 m) from the original data sets, so no additional calibration factors were applied. The profile with the total CO $J$=2–1 flux emitted by IRC +10420 is shown in Fig. 8. (For CO $J$=1–0 the total flux is shown in the profiles published by B01.)

The percentage of flux lost in the interferometric observations of AFGL 2343 is ∼ 30–40%, as deduced by comparison with the profiles obtained at PV (B01). When adding the zero-spacing B01 data to the PdB visibilities, the size of AFGL 2343 is found to be very similar at both frequencies and slightly smaller (but close) than the PV beam (at half power) at 1mm. This, and the comparison with the same analysis made for IRC +10420, leads us to conclude that the amount of flux lost in the PV profiles of AFGL 2343 (from B01) is probably quite small at 1mm, if there is any at all. Therefore, in order to recover the flux filtered out by the interferometer, we just added the strict zero-spacing data published by B01 to the PdB visibilities. For consistency with the flux calibration adopted from the OTF IRC +10420 observations, the flux of the CO $J$=1−0 and 2−1 PV profiles of AFGL 2343 have been increased by 6% and 9%, respectively.

The maps of CO $J$=2−1 and 1−0 from merging PV and PdB data are shown, respectively, in Figs. 2 and 3 for IRC +10420, and in Figs. 4 and 5 for AFGL 2343. The conversion factors from flux units to main-beam brightness temperature ($T_{\rm mb}$) units are, at 1.3 mm and 2.6 mm respectively, 12.0 K per Jy beam$^{-1}$ and 11.2 K per Jy beam$^{-1}$ for the IRC +10420 maps, and 11.0 K per Jy beam$^{-1}$ and 10.1 K per Jy beam$^{-1}$ for the AFGL 2343 maps. We resampled the velocity channels to a final velocity resolution of 5 km s$^{-1}$ for both transitions.

## 3. Analysis of the $^{12}$CO mapping; aspherical components

The main common property of the $^{12}$CO line profiles in both YHGs, IRC +10420 and AFGL 2343, is their large width, i.e. their CSEs are expanding at high velocities with respect to CSEs of typical AGB stars, ∼ 10-15 km s$^{-1}$(see e.g. Loup et al. 1993 and Ramstedt et al. 2006). The CO maps in Figs. 2–5 show that the envelopes of both YHGs are approximately spherical, so their circumstellar mass was ejected more or less isotropically; this is assumed in the modeling and discussed in Sect. 4. There are, however, some remarkable deviations from sphericity in the maps of both sources, which we analyze in more detail in the following sections.

### 3.1. IRC +10420

The CO maps in Figs. 2 and 3 show brightness distributions that are not completely circularly symmetric. First, the northern part of the envelope is considerably brighter than the southern regions, as can be seen in the central channels of the maps. Note that the southern less-bright region is very wide. To our knowledge, no aspherical structure as extended as this has been observed in the brightness distribution of AGB CSEs so far and have not been predicted by mass-loss models. Note that IRC +10420 is not an AGB star and that, for instance, the expansion velocity of its CSE (∼ 35 km s$^{-1}$) is much higher than for typical AGB stars.

Second, at velocities between 60 and 75 km s$^{-1}$, a bright component, which is elongated along an axis of PA ∼ −115° (see Fig. 6), is visible in the brightness distributions at 3 and 1mm. This elongation is possibly linked to the presence of a bipolar outflow, expanding along an axis close to the sky plane. There is no clear counterpart to this feature at red-shifted velocities, although a certain brightness increase is perceived close to the star in the northeastern direction at ∼ 100 km s$^{-1}$.

Third, in the innermost regions of the CO $J$=2−1 maps, in channels with LSR velocities from 70 km s$^{-1}$ to 95 km s$^{-1}$, we see two bright clumps located along an axis at PA ∼ 72°. They leave a relative brightness minimum at the CSE center, which suggests that there is a minimum in the mass distribution at the nebula center (see Sect. 4). The comparison of the position and shape of these bright knots with the synthetic beam suggests that a disk-like distribution may be embedded in the innermost

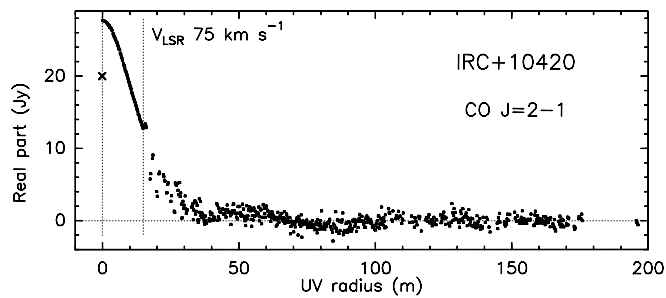

**Fig. 1.** $^{12}$CO $J$=2−1 visibilities (their real part vs. $uv$−radius) obtained for IRC +10420, at the LSR velocity of 75 km s$^{-1}$. A dotted vertical line at a $uv$−radius of 15 m separates the PdB data from the 30m OTF observations. The flux collected in the beam of the 30m telescope towards the center of IRC +10420 is shown at the zero-spacing, at $uv$−radius = 0 (marked with a cross).

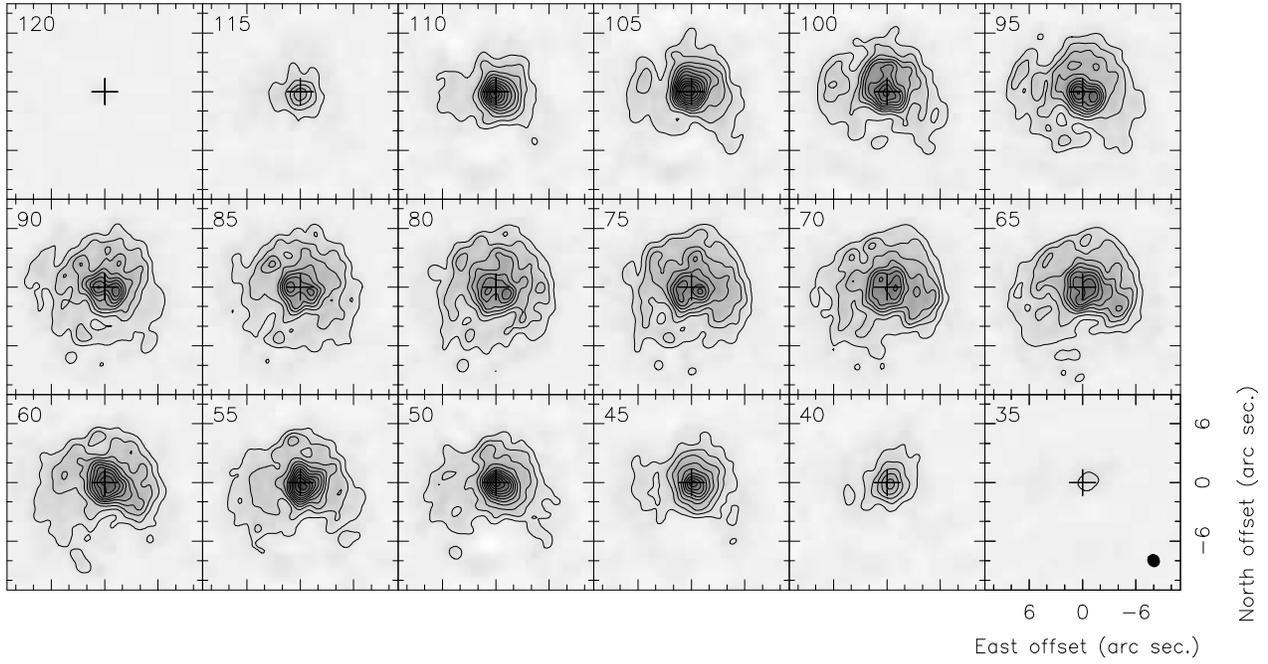

**Fig. 2.** Combined PdBI + 30m-telescope channel maps of the $^{12}$CO $J$=2–1 emission in IRC +10420 (J2000 central coordinates: $19^h26^m48\overset{s}{.}10$, +11°21′17″.0). LSR velocities (in km s$^{-1}$) are indicated in the upper left corner of each box. The first contour and level step are at $11\times\sigma$ = 0.14 Jy/beam. There are no negative contours at −0.14 Jy/beam. The CLEANed beam (at half-power level), of size 1″.5×1″.3 (FWHM) and position angle (PA) 120°, is drawn in the bottom right corner of the last panel.

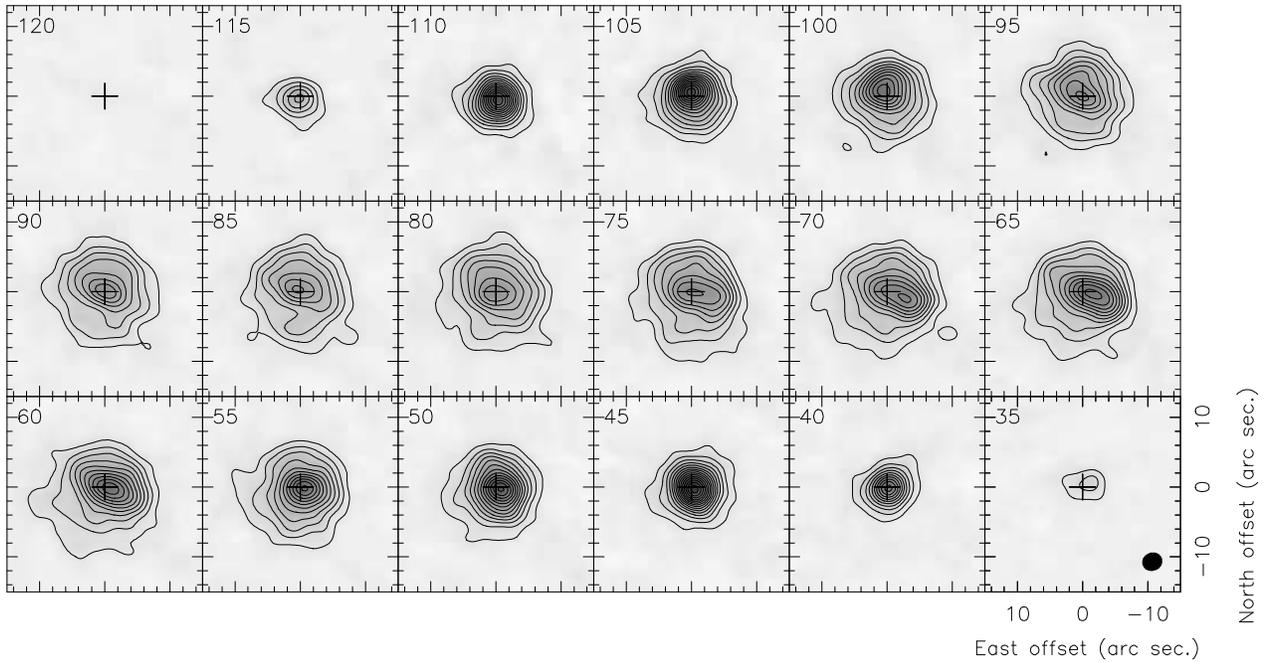

**Fig. 3.** Combined PdBI + 30m-telescope channel maps of the $^{12}$CO $J$=1–0 emission in IRC +10420 (same central coordinates as in Fig. 2). LSR velocities (in km s$^{-1}$) are indicated in the upper left corner of each box. The first contour and level step are at $5\times\sigma$ = 0.05 Jy/beam. There are no negative contours at −0.05 Jy/beam. The CLEANed beam (at half-power level), of size 3″.1×2″.6 (FWHM) and PA 77°, is drawn in the bottom right corner of the last panel.

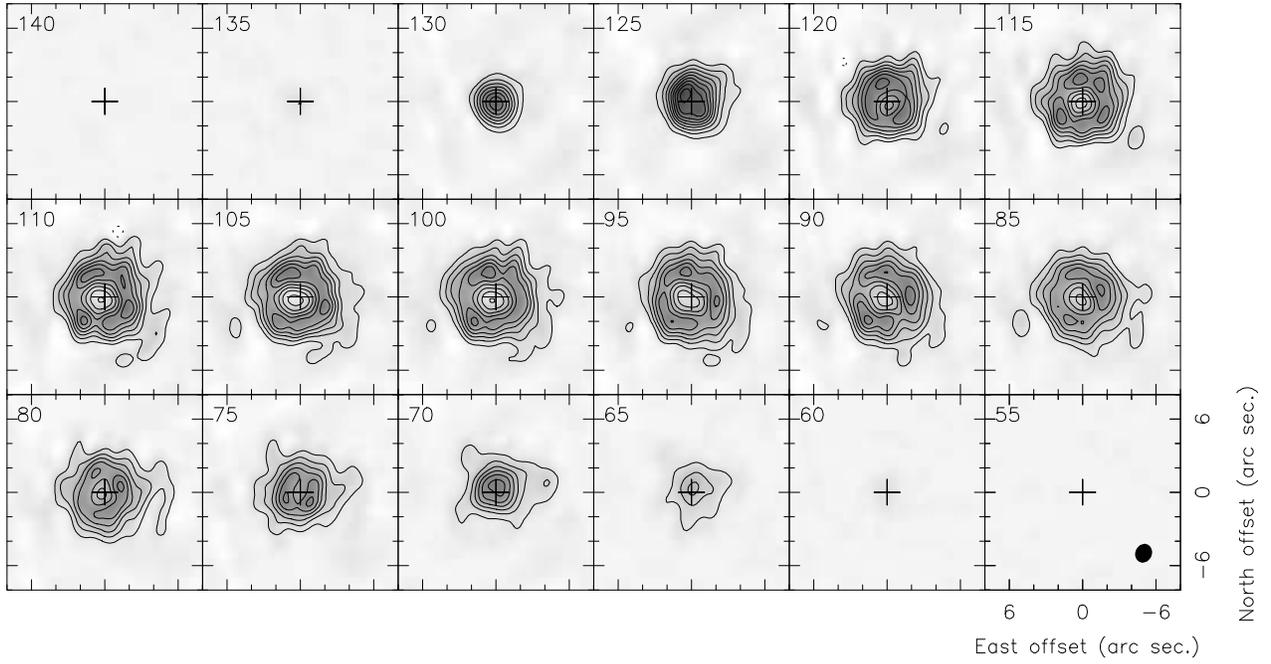

**Fig. 4.** PdBI (+ zero-spacing) maps of the $^{12}$CO $J$=2−1 emission in AFGL 2343 (J2000 central coordinates: $19^h13^m58\overset{s}{.}60$, +00°07′32″.00). LSR velocities (in km s$^{-1}$) are indicated in the upper left corner of each box. The first contour and level step are at 14×$\sigma$ = 0.2 Jy/beam. A negative level at −0.2 Jy/beam is shown in dashed contours. The CLEANed beam (at half-power level), of size 1″.5×1″.3 (FWHM) and PA 158°, is drawn in the bottom right corner of the last panel.

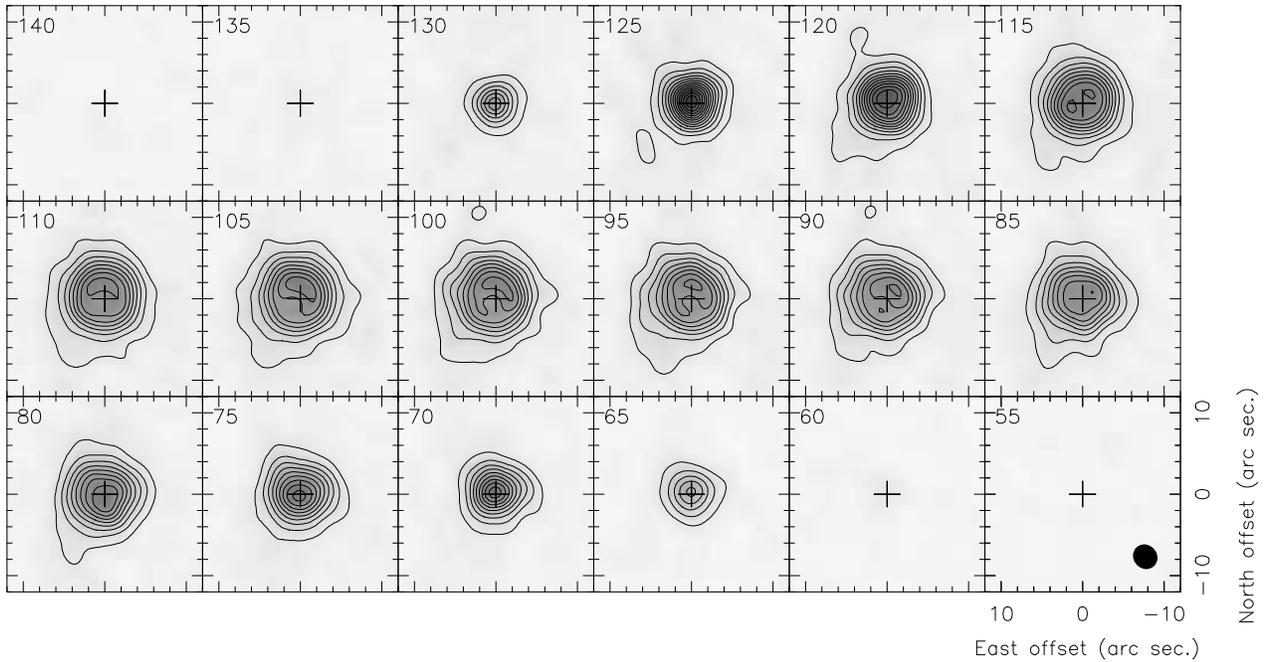

**Fig. 5.** PdBI (+ zero-spacing) maps of the $^{12}$CO $J$=1−0 emission in AFGL 2343 (same central coordinates as in Fig. 4). LSR velocities (in km s$^{-1}$) are indicated in the upper left corner of each box. The first contour and level step are at 11×$\sigma$ = 0.12 Jy/beam. There are no negative contours at −0.12 Jy/beam. The CLEANed beam (at half-power level), of size 3″.1×2″.9 (FWHM) and PA 42°, is drawn in the bottom right corner of the last panel.

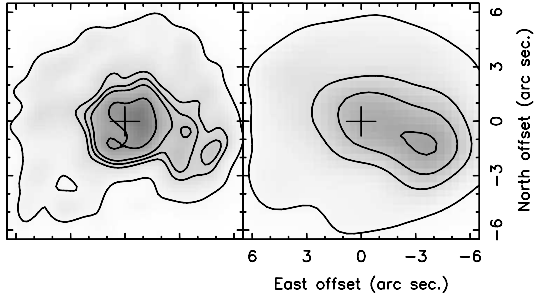

**Fig. 6.** The average of the brightness at 65 and 70 km s$^{-1}$ for CO $J$=2−1 (from Fig. 2; here on the left) and CO $J$=1−0 (from Fig. 3; here on the right) for IRC +10420. In order to emphasize the presence of a bright clump elongated towards the southwest direction, we have plotted contours at 0.2, 0.5, 0.6, 0.7, and 0.96 Jy/beam for CO $J$=2−1 and at 0.1, 0.25, 0.35, and 0.45 Jy/beam for CO $J$=1−0.

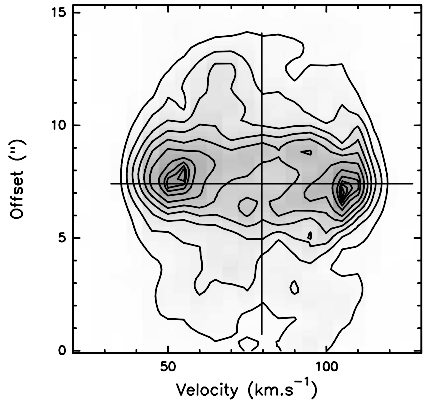

**Fig. 7.** CO $J$=2−1 position-velocity diagram of IRC +10420 along an axis at PA 72°, passing through both innermost brightness peaks. Offsets from 0 to 15″ in the vertical axis correspond to offsets in the axis at PA 72° from east to west in the maps in Fig. 2. Contours are plotted every 0.2 Jy/beam, from 0.2 to 1.4 Jy/beam, and at the intensity levels of 1.5, 1.55, 1.57, and 1.6 Jy/beam.

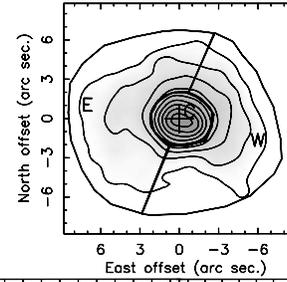

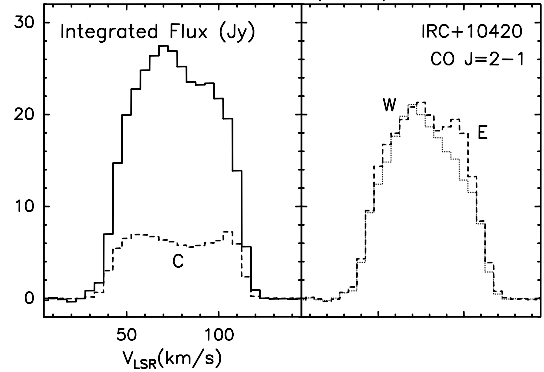

**Fig. 8.** *Top*: Averaged CO $J$=2−1 emission, divided by a solid line in three regions, labeled as C, E, and W. *Bottom*: On the left, the profile obtained from integrating the whole nebular emission (solid line) and the total emission from the C-labeled region (dashed line). On the right, the flux coming from the E- (dashed line) and W-labeled (dotted line) regions.

shell. The knots do not seem to be due to the convolution of a shell distribution with the elongated synthetic beam since they are not aligned with it. In Fig. 7 we have plotted a position-velocity diagram along the axis at PA 72°, passing through both clumps. A symmetrical pattern is observed for the two brightest features found in this diagram, which are separated up to 0.″4 from the center in opposite directions. Compatible features are observed for the same position-velocity diagram in CO $J$=1−0 maps. Given their high relative velocities (with respect to the systemic velocity, V$_*$), we cannot propose Keplerian rotation for this innermost gas, since this would require too high a central mass (∼ 1000$M_\odot$). An expanding disk (or torus) embedded in the innermost shell or/and some oblateness in the CSE could explain such a structure and seem a more reasonable explanation.

Finally, we investigated the asymmetry observed in the CO profiles (see Fig. 8). The extra emission responsible for the non-centered profile peak does not result from the bright elongation observed between 60-70 km s$^{-1}$ (see Fig. 6), but it comes from the whole CSE. This asymmetry seems to come from a lack of emission at velocities higher than 70 km s$^{-1}$, mainly from the westernmost (W-labeled) hemisphere (see Fig. 8).

### 3.2. AFGL 2343

The CO maps in AFGL 2343 (Figs. 4 and 5) present a shell-like brightness distribution that is quite uniform for a given radius. The central velocity maps at 1mm, however, show the presence of clumpiness. Clumps were also observed, e.g., in the detached shells around the AGB stars S Sct and TT Cyg (Olofsson et al. 1992, 2000). Several reasons have been brought up to explain clumpiness in CSEs of AGB stars (i.e. Woitke et al. 2000; Schirrmacher et al. 2003; Socker 2002; Bergman et al. 1993), although they have not yet been parametrized into mass-loss models to reproduce observations.

In addition to the clumps, although perhaps related to them, there is an increase in the brightness distribution for both transitions in the upper part of the shell (see Fig. 5). By convolving the 1mm maps with the synthetic beam obtained at 3mm, we obtain CO $J$=2−1 maps that are very compatible with those observed in CO $J$=1−0.

Finally, at the lowest-intensity contours, a marginal elongation is detected in the CO $J$=1−0 maps in the southeast direction. A wider dynamic range and perhaps proper short-spacing observations would allow this to be confirmed.

## 4. Nebula emission model

We used a radiative transfer code to model the CO $J$=2−1 and 1−0 emission of the two YHGs: AFGL 2343 and IRC +10420.

High-$J$ CO transitions (from Teyssier et al. 2006) were also taken into account in the modeling of IRC +10420. In view of the more or less spherical, though not always very regular, shapes of the CO emission distributions, we assumed spherical symmetry and isotropic expansion. Line excitation and level populations are calculated using a standard LVG approach. Radiative excitation considers the IR emission by the star and the innermost circumstellar dust through the CO $\Delta v$=1 vibrational transitions at 4.7 $\mu$m, following the method already described by Teyssier et al. (2006). The IR emission is assumed to be that of a black body whose total intensity is adjusted to yield the observed flux at this wavelength; we note that the assumed frequency dependence has a negligible effect on the calculations, since the relative variation in the frequency of the different rovibrational components is very small. Collisional excitation is also accounted for using coefficients calculated by Flower (2001) and an abundance of ortho-$H_2$ three times that of para-$H_2$. We also took into account the extrapolation to higher values of the temperature and $J$-numbers in the *Leiden Atomic and Molecular Database*[2] (LAMDA; see Schöier et al. 2005). We note that this extrapolation has little effect on our calculations, in any case, because the temperatures in our sources are moderate, only exceeding 400 K in the innermost layers around IRC +10420. The level populations so determined are used to derive emission and absorption coefficients. The brightness distribution is calculated by solving the standard radiative transfer equations and is later convolved with a Gaussian beam, yielding main-beam Rayleigh-Jeans-equivalent temperatures, $T_{\rm mb}$, directly comparable to the observations. In this process, we assume both a macroscopic velocity and a local velocity dispersion (due to thermal or turbulent movements).

In the model, the density distribution is given by an isotropic mass-loss rate ($\dot{M}$) and the expansion velocity ($V_{\rm exp}$). The temperature at a given radius $r$ is described by a potential law, as usually assumed for AGB circumstellar envelopes: $T_{\rm k}(r) = T(r_{\rm o}) \times (r/r_o)^{-\alpha_t} + T_{\rm min}$. The thermodynamics of the gas around YHGs is not well known; potential temperature laws have been shown to be a reasonable approximation for AGB circumstellar envelopes and are known to be compatible with CO observations in such sources (e.g. Schöier & Olofsson 2001; Teyssier et al. 2006, and references therein). The local velocity dispersion is described assuming a Gaussian dispersion given by a standard deviation $\sigma_{\rm turb}$. All these parameters are assumed to be constant within finite shells, and sequences of shells with different parameter values are introduced to reproduce variations in the mass-loss history. The inner and outer radii of the different shells ($R_{\rm in}$, $R_{\rm out}$), the logarithmic velocity gradient ($\epsilon$), and the relative CO abundance ($X_{\rm CO}$) are other input parameters in our code. $X_{\rm CO}$ is assumed to be 3 $10^{-4}$ and constant in the envelopes. The code and approximations are described in more detail by Teyssier et al. (2006).

### 4.1. Model-fitting results

Since we are assuming spherical symmetry, we have fitted with our model the azimuthal average of the brightness distributions shown in Sect. 2, $T_{\rm mb}(r)$. Note, however, that IRC +10420 in particular shows significant departures from circular symmetry (see Figs. 2 and 3, and Sect. 3.1). In modeling IRC +10420 we assumed an uncertainty of 10% in the relative flux calibration of both lines.

[2] http://www.strw.leidenuniv.nl/~moldata/

Model predictions that simultaneously fit the $T_{\rm mb}(r)$ of both transitions, CO $J$=1−0 and 2−1, are derived for several LSR velocities. Most of our results are deduced from modeling at the central velocities, in which the structure of the envelope is most obvious. Nevertheless, fitting data at extreme velocities provides crucial information on the velocity fields in the different nebular components.

The HPBW used in the modeling of each map is a circular approximation (with the same area) of the slightly-elliptical synthetic beams obtained in the CLEANing process (see Sect. 2). For IRC +10420, the HPBW of the assumed circular beam is 2″.86 and 1″.39 for CO $J$=1−0 and 2−1, respectively. For AFGL 2343, it is 3″.02 for CO $J$=1−0 and 1″.44 for CO $J$=2−1.

The parameters of the model best-fitting our data are shown in Tables 1 and 2. Figures 9 and 10 show the observations compared with model predictions for the central velocity channels and for both CO transitions, in AFGL 2343 and IRC +10420 respectively, on the left. The mass-loss history deduced for each source is shown on the right.

In AFGL 2343 we have identified several shells (see Fig. 9) in our model. They correspond to a main mass-loss phase, followed in time by a very sharp mass-loss decrease. A very high mass-loss rate is found in the main shell, shell 2 in Table 1, which is likely what is responsible for the formation of the ring-like structure shown in the CO $J$=2−1 map. The emission in this dense region is opaque, and therefore the deduced mass-loss rate is probably just a lower limit. Shell 3 can be seen as a connection between shells 2 and 4, showing the increase in the mass-loss rate. Low temperatures are found, especially in the densest regions of the envelope. $\alpha_t$ is deduced to be low, denoting a slow variation in the temperature with radius.

The envelope of IRC +10420 is less dense than that of AFGL 2343. The temperature in the innermost regions is higher than in AFGL 2343, but in the outermost regions it becomes similar due to the higher value derived for $\alpha_t$. The mass-loss rate found is not as high as in the previous source, but is high compared with standard evolved nebulae. Two different periods of mass loss are needed to fit the data, separated by a short phase of much lower mass loss. We have mentioned that the circumstellar envelope in IRC +10420 is roughly circular on a large scale, but shows many clumps and a significant substructure. Our fitting with a spherically symmetric model certainly helps in getting an idea of the mass-loss history of the source, but is obviously too simplistic to reproduce the details of the brightness distribution. See Fig. 11 and discussions in Sects. 3.1 and 4.2.

The main problem we have found in our fitting is in how the source size varies with the observed velocity ($V_{\rm obs}$). The source size is predicted for each spherical elementary shell, with characteristic radius $R_s$, and must yield an increase in brightness for velocities close to the extreme ones. We have to take into account that the observed radius of a shell varies with $V_{\rm obs}$:

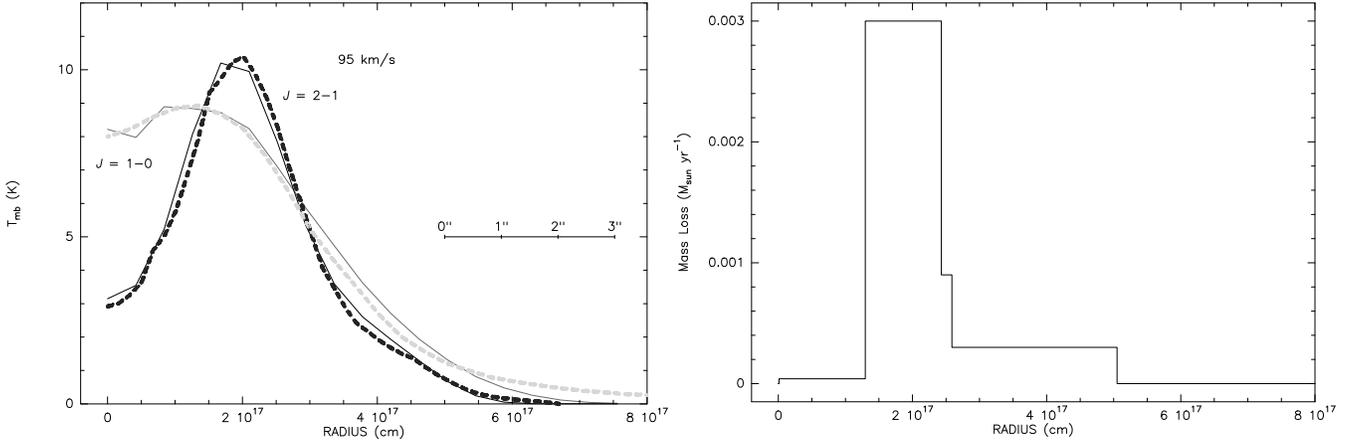

**Fig. 9.** *Left:* Azimuth-averaged brightness distribution (in dashed lines) compared with model results (solid lines) for AFGL 2343, at $V_{\rm LSR} = 95$ km s$^{-1}$. CO $J$=1–0 data are plotted in grey, $J$=2–1 data in black. *Right:* Mass-loss pattern found for AFGL 2343.

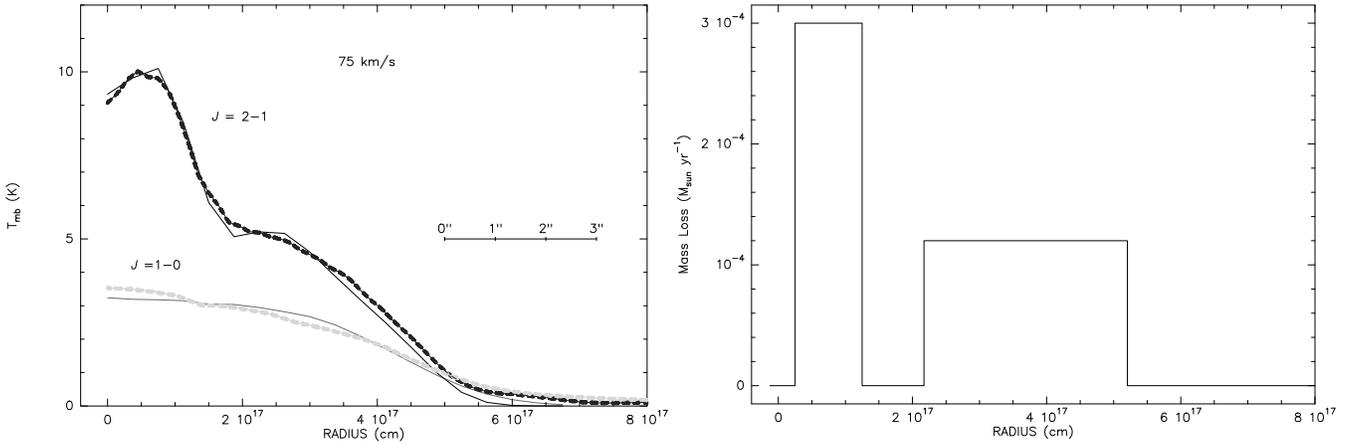

**Fig. 10.** *Left:* Azimuth-averaged brightness distribution (dashed lines) compared with the model results (solid lines) for IRC +10420, at $V_{\rm LSR} = 75$ km s$^{-1}$. CO $J$=1–0 data are plotted in grey, $J$=2–1 data in black. *Right:* Mass-loss pattern found for IRC +10420.

**Table 1.** Parameters of the best fit for AFGL 2343.

| Shell | $R_{\rm in}$ (cm) | $R_{\rm out}$ (cm) | $\dot{M}(M_\odot$ yr$^{-1})$ | $T(r_{\rm o} = 10^{17}$cm) (K) | $\alpha_t$ | $V_{\rm exp}$(km s$^{-1}$) | $\sigma_{\rm turb}$(km s$^{-1}$) | $\epsilon$ |
|---|---|---|---|---|---|---|---|---|
| 1 | 1 10$^{15}$ | 1.3 10$^{17}$ | 4 10$^{-5}$ | 29 | 0.5 | 35 | 3 | 0.1 |
| 2 | 1.3 10$^{17}$ | 2.5 10$^{17}$ | 3 10$^{-3}$ | 13 | 0.5 | 35 | 3 | 0.1 |
| 3 | 2.5 10$^{17}$ | 2.6 10$^{17}$ | 9 10$^{-4}$ | 21 | 0.7 | 35 | 3 | 0.1 |
| 4 | 2.6 10$^{17}$ | 5 10$^{17}$ | 3 10$^{-4}$ | 40 | 0.7 | 35 | 3 | 0.1 |

$T_{\rm min} = 8$ K remains the same for all the assumed shells. (See the description of the parameters in Sect. 4.)

**Table 2.** Parameters of the best fit for IRC +10420.

| Shell | $R_{\rm in}$ (cm) | $R_{\rm out}$(cm) | $\dot{M}(M_\odot$ yr$^{-1})$ | $T(r_{\rm o} = 10^{17}$cm) (K) | $\alpha_t$ | $V_{\rm exp}$(km s$^{-1}$) | $\sigma_{\rm turb}$(km s$^{-1}$) | $\epsilon$ |
|---|---|---|---|---|---|---|---|---|
| 1 | 2.5 10$^{16}$ | 1.24 10$^{17}$ | 3 10$^{-4}$ | 230 | 1.2 | 37 | 3 | 0.1 |
| 2 | 2.2 10$^{17}$ | 5.2 10$^{17}$ | 1.2 10$^{-4}$ | 100 | 0.8 | 25 | 20 | 0.3 |

$T_{\rm min} = 3$ K remains the same throughout the whole envelope. (See the description of the parameters in Sect. 4.)

$R(V_{\rm obs}) = R_s \sqrt{1 - \frac{(V_{\rm obs}-V_*)^2}{V_{\rm exp}^2}}$. In a resolved CSE, the ratio between $T_{\rm mb}(V_{\rm obs} \neq V_*)$ and $T_{\rm mb}(V_{\rm obs} = V_*)$ must be at its maximum for optically thin emission and tends to 1 for optically thick lines. In the azimuth-averaged brightness of IRC +10420, this ratio is, however, significantly lower than 1 for several values of $R_s$ (within the outer shell), $R$ and $V_{\rm obs}$ (related by the above formula), so our model cannot reproduce the emission for such high velocities. We think that this is due to the fact that the coherence lengths[3] for extreme velocities in the actual envelope are smaller than the large lengths expected under spherical symmetry, due to the high clumpiness character-istic of IRC +10420. Such a phenomenon can only be introduced in our model assuming a very high value of $\sigma_{turb}$. We can reproduce the observed variations in the brightness distributions with $V_{obs}$ if we take $\sigma_{turb} \sim 20$ km s$^{-1}$ for the outer shell of IRC +10420 (while the expansion velocity in this shell decreases to about 25 km s$^{-1}$). We are aware that this probably does not correspond to the real situation, but is more related to clumpiness, and so our assumption on $\sigma_{turb}$ could just serve to check that our hypothesis about the limits to the actual coherence lengths is likely to be true. Note that the fact that the value of $\sigma_{turb}$ in our model for the second shell is comparable to the expansion velocity can be translated into the coherence length being severely limited by clumpiness in the actual envelope for all $V_{obs}$.

---

[3] In a given direction, the distance in which the change of the projected macroscopic velocity is small enough to allow radiative interaction.

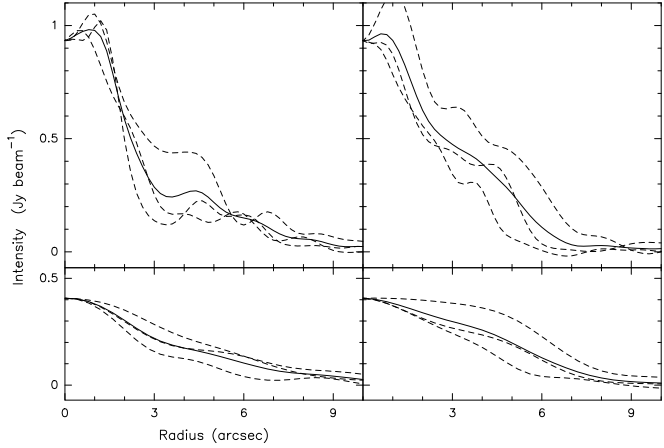

**Fig. 11.** Brightness distributions of the CO $J$=2–1 (*upper*) and 1–0 (*bottom*) emissions in IRC +10420 (from maps in Figs. 2 and 3) along six axes, starting from the nebula center and equidistant. On the *left*, in dashed lines, data for the northwestern axes at PA 5°, 305°, and 245° (coinciding with the elongation emphasized in Fig. 6). On the *right*, in dashed lines, data for the southeastern axes at PA 65°, 125°, and 185°. Averages of the data shown in each plot in dashed lines are presented in solid lines.

The outermost radius derived for both envelopes, AFGL 2343 and IRC +10420, is almost the same and the time scales of the mass-loss phases are similar. The innermost shell of IRC +10420 is comparable to the one detected in SiO emission by Castro-Carrizo et al. (2001a). This SiO shell was found to extend from 0.9 10$^{17}$cm to 1.4 10$^{17}$cm, which is in the outer zone of the innermost CO shell. This suggests that the SiO-rich shell could correspond to a shock front associated to the formation of a densest shell traced by CO.

### 4.2. Uncertainties in the fitting process

In a complex fitting process like the one performed here, it is difficult to estimate the uncertainties in the values derived for the parameters (implicitly or explicitly) involved in the model. Some of them are, however, quite directly deduced from the observations and are therefore as reliable as the data itself. This is the case, in particular, for the velocity and extent of the shells (at least of the main components), given by the total velocity dispersion and extent of the detected features, respectively. Therefore, the characteristic times are also quite reliable.

Uncertainties due to the sphericity assumption exist for IRC +10420 and AFGL 2343, IRC +10420 showing the most significant departures from symmetry. As an example, we present the brightness distribution of CO $J$=2–1 and 1–0 along axes in several directions, crossing the nebula center, in Fig. 11. We see there that the maximum of the emission in all cuts appears at a distance of about 0″.5 – 1″, with a relative minimum at the center. In our opinion, this confirms a very significant decrease in the mass-loss rate at present, as derived from our fitting of the azimuth-averaged brightness distribution. We also note that in all the directions there is a secondary maximum or hump at distances of about 4″ – 5″, a relative minimum often appearing at about 3″. This feature appears quite irregular and significantly varies between the different cuts. It confirms that the outermost regions of the envelope are strongly aspherical, but shows that a significant amount of material is placed at such a distance from the star, being probably detached from the intense central component. This material corresponds to the shell 2 deduced from our fitting. The differences in the outermost ($\sim$3″–6″) brightness distribution for the different directions must be kept in mind to evaluate the uncertainties on the properties deduced for the outermost detached layer: its existence is very probable and its total mass is likely to be close to that given by our fitting, but its structure remains poorly known. Finally, we notice that uncertainties from the quality of the fittings are negligible in comparison with those resulting from the isotropy assumption. In order to show this, we estimated the standard deviation of the azimuthal T$_{mb}$ average at each radius. For distances from the star of 1, 2.5, 4, and 6 × 10$^{17}$ cm, respectively, we obtain for the average of the AFGL 2343 CO 2–1 T$_{mb}$ the rms (in K) 1.3, 2.3, 1.0, and 0.4; for the AFGL 2343 CO 1–0 T$_{mb}$ the rms 0.5, 1.2, 0.8, and 0.4; for the IRC +10420 CO 2–1 T$_{mb}$ the rms 1.7, 2.0, 1.2, and 0.4; for the IRC +10420 CO 1–0 T$_{mb}$ the rms 0.5, 0.7, 0.6, and 0.2. Taking these figures into account, the actual presence and characteristics of shell 3 in AFGL 2343 are uncertain. Shell 1 is somehow needed in AFGL 2343, although its properties can also be questioned.

We have also discussed in Sect. 4.1 the uncertainties in the velocity field of the second shell of IRC +10420 due to the lack of spherical symmetry.

The mass-loss rates and total envelope mass are affected by two important factors. 1) The moderately high optical depth is approximately treated by the LVG model but yields a low dependence of the intensity on the number of emitting molecules. As usual, the main uncertainty introduced by the optical depth is a possible underestimate of the total mass. 2) The abundance of $^{12}$CO is poorly known. We have assumed a high value for the relative abundance, similar to what is often measured in O-rich AGB shells, but if this molecule is less abundant by some factor (due perhaps to dissociation), our mass values would also be underestimated by the same factor.

The masses derived here are 1.2 (for AFGL 2343) and 2 (for IRC +10420) times lower than those measured from $^{13}$CO $J$=1–0 by Bujarrabal et al. (2001). This line is very probably optically thin, but we note that the abundance of $^{13}$CO is also not well known and that the treatment of the excitation in that paper is relatively simple.

Our mass values in AFGL 2343 are 1.5 - 2 times lower than the total masses derived from model fitting of the dust emission (Buemi et al. 2006; Hawkins et al. 1995; Gledhill et al. 2002). In IRC +10420, the comparison is more difficult because in most cases only mass-loss rates have been calculated from dust emission observations; in any case, the rates calculated from IR continuum data are also about a factor 2 higher than ours (Hrivnak et al. 1989; Oudmaijer et al. 1996). This discrepancy could be due to an underestimate of the total mass in our calculations or to the existence of a high amount material at distances $\gtrsim 10^{18}$ cm, poor in molecules due to photodissociation by the interstellar ultraviolet field, as discussed below (among other sources of error in such a comparison, like an unexpected high abundance of grains).

The outer photodissociation radii expected in our sources, due to the interstellar UV field, are larger than the total radii found here. For a shell with an expansion velocity of $\sim$ 35 km s$^{-1}$, a CO relative abundance equal to 3 10$^{-4}$, and mass-loss rates of 3 10$^{-4}$ and 1.2 10$^{-4}$ $M_\odot$ yr$^{-1}$ (corresponding to the outer shells of AFGL 2343 and IRC +10420, respectively), the photodissociation theory (Mamon et al. 1988) predicts outer radii of, respectively, $10^{18}$ cm and 6.5 10$^{17}$ cm. These values are significantly higher than those derived from our data, mainly for AFGL 2343, and much higher than the radii of the most massive inner shells, suggesting that the shell radii derived here represent a real decrease in the mass-loss rate and not photodissociation of CO. This conclusion is supported by the fact that the extent of the envelope around IRC +10420 is larger towards the north or northwest, more or less in the direction of the galactic plane, from which we would expect the largest UV radiation. The photodissociation radii are not, however, much larger than the outermost radii, so we cannot exclude that other outer shells poor in molecules have not been detected but instead contribute to the total circumstellar mass.

The innermost radii measured here are not expected to be due to CO photodissociation by the stellar radiation, because of the cool central stars. ISO observations of atomic lines characteristic of PDRs (Castro-Carrizo et al. 2001b; Fong et al. 2001) did not yield detections of PDRs around evolved stars with surface temperatures lower than $10^4$ K. In those nebulae, the PDR mass limits were usually a tiny fraction of the total molecular mass. Our stars are cooler than $10^4$ K, particularly IRC +10420. Castro-Carrizo et al. (2001b) observed AFGL 2343 and found no trace of atomic FIR emission. Images of dust emission (Sect. 1) support the absence of circumstellar material in the innermost regions.

It is difficult to determine the temperature law from fitting just the observed lines, corresponding to low-energy levels. Only in the outermost shells, where the temperatures are $\lesssim$ 20 K, the fitting of the 1–0 and 2–1 lines is relevant at this respect. (On the other hand, this implies that the assumed temperature law does not strongly affect the mass-loss rate estimates.) To improve the temperature estimate, we also tried to reproduce high-$J$ CO lines from Teyssier et al. (2006). The main result concerns the inner shell of IRC +10420. The high $J$=6–5 line intensity observed in this source requires a high temperature in the innermost shell, from which this line mainly emits, higher than $\sim$ 200 K. But we note that if the empirical datum is over-calibrated by more than a factor 1.5 (which is improbable but not impossible at this frequency, see discussion by Teyssier et al.), a temperature law that is more similar to the one found in the other shells could be acceptable. In AFGL 2343, where high-$J$ observations just include upper limits, we have checked that our model is compatible with these data. In fact, the low emission at high frequency of this source seems to be due to the empty inner region.

An important result found here is the existence of intervals of time in which the mass-loss rate of both sources has been very low, close to zero. We checked that significant mass-loss rates strongly affect the model predictions, assuming that the rest of the parameters do not vary. For the central, almost empty region in both stars, we find that the mass-loss rate must be lower than $\sim$ 5 10$^{-5}$ $M_\odot$ yr$^{-1}$ to keep the predictions compatible with our maps. A limit $\sim$ 10$^{-5}$ $M_\odot$ yr$^{-1}$ is found for the interface region in IRC +10420, at about 2 10$^{17}$ cm from the star, but in this case we recall the uncertainties due to asphericity in the extended layers. We note that all these values are very low compared with the mass-loss rates found in the adjacent regions.

## 5. Evolution of the circumstellar envelopes around YHGs

### 5.1. Kinematical times

Taking the radii of each shell and the deduced expansion velocity into account (Tables 1 and 2), we estimate the duration of each mass-loss period (see Table 3).

From the kinematical times derived for AFGL 2343 (see Table 3), we find that this source had a mass-loss event starting $\sim$ 4500 yr ago. After $\sim$ 2200 yr of a high mass-loss rate (3–9 10$^{-4}$ $M_\odot$ yr$^{-1}$), it increased to an extremely high value ($\sim$ 3 10$^{-3}$ $M_\odot$ yr$^{-1}$) that was kept constant for $\sim$ 1000 yr. After that period the amount of mass expelled to the interstellar medium strongly decreased and remained constant ($\sim$ 4 10$^{-5}$ $M_\odot$ yr$^{-1}$) during the last $\sim$ 1200 yr.

**Table 3.** Kinematical times deduced for each mass-loss period for AFGL 2343 and IRC +10420.

| AFGL 2343 | | | IRC +10420 | | |
|---|---|---|---|---|---|
| Shell* | $\dot{M}$ | t | Shell* | $\dot{M}$ | t |
| | ($M_\odot$ yr$^{-1}$) | (yr) | | ($M_\odot$ yr$^{-1}$) | (yr) |
| 1 | 4 10$^{-5}$ | 1200 | | 0 | 200 |
| 2 | 3 10$^{-3}$ | 1100 | 1 | 3 10$^{-4}$ | 800 |
| 3 | 9 10$^{-4}$ | 90 | | 0 | 1200 |
| 4 | 3 10$^{-4}$ | 2100 | 2 | 1.2 10$^{-4}$ | 3800 |

* Shells with larger reference numbers were formed before.

The mass-loss pattern of IRC +10420, shown in Fig. 10, reveals the existence of two (probably separated) shells formed in the past $\sim$ 6000 yr. One shell is wider and farther away from the star, and the other is narrower and closer to it. The first mass-loss episode (1.2 10$^{-4}$ $M_\odot$ yr$^{-1}$) lasted $\sim$ 3800 yr and is responsible for the extended component found in the CO maps. After this time, the mass loss (almost) stopped for around 1200 yr. Another important mass-loss period (3 10$^{-4}$ $M_\odot$ yr$^{-1}$) then started, lasting ~ 800 yr and resulting in the innermost shell, apparently very close to the region where SiO is detected. The mass loss seems to have stopped in the past ~ 200 yr.

### 5.2. Circumstellar mass in YHGs

The total masses deduced for the circumstellar envelopes of these YHGs are ~ 4 $M_\odot$ for AFGL 2343 and ~ 1 $M_\odot$ for IRC +10420. Note that these values are lower by a factor 1.2 and 2 than those obtained by B01 from data of the two first rotational transitions of $^{13}$CO (Sect. 4.2). This difference may be due to opacity effects, which affect $^{12}$CO more than any other isotopic species. It is also possible, however, that these massive O-rich sources present a particularly high $^{13}$C abundance. Our mass values are also significantly lower than those obtained from dust emission analysis; see further discussion in Sect. 4.2.

In order to compare the mass of the circumstellar environment characteristic of YHGs, we used the IR emission at 60 $\mu m$ from the IRAS catalog; see Table 4. In this table, we can see that IRC +10420 and AFGL 2343 are by far the YHGs with the highest infrared emission. These two sources show very massive circumstellar envelopes, probably rare among YHGs.

**Table 4.** Infrared flux (F) at a wavelength of 60$\mu m$ for YHGs at distances $D^*$

| source | other name | $F_{60\mu m}$(Jy) | $D^*$(kpc) |
|---|---|---|---|
| IRC +10420 | - | 718 | 5.0 |
| AFGL 2343 | HD 179821 | 516 | 5.6 |
| HD 119796 | HR 5171 A | 127 | 3.6 |
| KY Cyg | IRC +40415 | 50.7 | – |
| MWC 300 | IC 6-23 | 41.1 | 0.5 |
| HD 212466 | RW Cep | 27.4 | 0.8 |
| HD 80077 | - | 3.74 | 1.25 |
| HD 96918 | V382 Car | 1.10 | 1.8 |
| HD 224014 | $\rho$ Cas | 0.912 | 3.6 |
| HD 74180 | HR 3445 | 0.717 | 0.95 |
| HD 223385 | 6 Cas | 0.517 | 5 |
| HD 217476 | HR 8751 | 0.463 | 6.7 |

$^*$ Distances derived from Hipparcos data except for IRC +10420 (from de Jager, 1998).

The expansion velocities deduced for AFGL 2343 and IRC +10420 are ~ 35 km s$^{-1}$, which are much higher than those usually found in AGB circumstellar envelopes. Such high velocities are also found in planetary and protoplanetary nebulae, but are related to fast bipolar flows carrying much less mass than the envelopes around YHGs.

In order to find out if such fast and massive flows can be driven by radiation pressure, the $P/(L/c)$ relation is used, where $P$ is the linear momentum (in fact a scalar magnitude, see discussion in e.g. B01) and $L$ is the stellar luminosity. The relation $P/(L/c)$(yr) must be comparable to the time during which the envelopes were formed. Since the momentum depends on the mass of the envelope, we take the mass deduced from the $^{13}$CO emission into account; luminosities are taken from B01. The values found range between $P/(L/c) \sim 2\ 10^3$ yr and ~ 10$^4$ yr. Taking the effects of multiple scattering by dust grains into account, which could increase the pressure acting onto grains, these times are compatible with the times deduced for the formation of these envelopes, which are (see Sect. 5.1) ~ 2 – 4 10$^3$ yr. The fast massive outflows detected in IRC +10420 and AFGL 2343 can therefore be driven by radiation pressure, contrary to what is obtained for fast post-AGB outflows.

### 5.3. Discussion; mass loss in YHGs

A similar episodic mass loss is found in both YHGs. Both sources show periods of enhanced mass loss separated by phases with a low or even negligible mass-loss rate, resulting in the detached shells observed. The characteristic times of the phases of mass loss are a few thousand years for both envelopes. Such events may be part of the evolution of YHGs.

The YHGs are supposed to find an instability region during their bluewards evolution called the *yellow void* (Niewenhuijzen & de Jager 1995). In the boundary of this region, the value of the effective gravity force, $g_{eff}$, approaches zero and becomes negative as the star gets into the instability region. When a pulsating hypergiant reaches the yellow void, stellar pulses can be strongly enhanced resulting in an episode of enhanced mass loss. As the star loses mass, its effective temperature decreases, so it moves redwards in the HR diagram, leaving the yellow void and ending the mass loss. Then, the temperature starts to increase moving the star bluewards again and, finally, it comes in contact with the yellow void once more. This process is called 'bouncing against the yellow void'.

A short-period bouncing against the void has been proposed by de Jager (1998) for the hypergiant HR 8752. For this source, two periods of mass loss were found within ~ 30 yr from the broadening of IR lines (Smoliński et al. 1989). The mass of circumstellar material found for this source is quite low (Smoliński et al. 1977), so the mass-loss processes presented by this source are much weaker than in AFGL 2343 and IRC +10420. If the circumstellar shells found here, which are exceptionally massive compared to those in other YHGs (see Sect. 5.2 and Table 4), are the result of a bouncing against the yellow void, AFGL 2343 and IRC +10420 should enter deep in the instability region and reach a negative enough value of $g_{eff}$ to drive to the very high mass-loss rates found. This large ejection of material would result in a large decrease in $T_{eff}$. These wide movements in the HR diagram could explain the long time scales found in the mass-loss episodes in our sources.

The rarity of the presence of massive envelopes in the YHGs might be due to an exceptional evolutionary path followed by AFGL 2343 and IRC +10420, like a long period bouncing, or to being at a different evolutionary status from the other hypergiants.

### 6. Conclusions

1. We mapped the emission of the transitions $^{12}$CO $J$=1–0 and $J$=2–1 in the yellow hypergiants (YHGs) AFGL 2343

and IRC +10420. Observations were performed with the Plateau de Bure interferometer. Data were also obtained with the 30m telescope in order to recover the flux filtered out by the array.

2. Our maps present a detached shell in AFGL 2343, showing a noticeable clumpiness. For IRC +10420, an inner detached shell was also detected, which approximately corresponds to the SiO emitting shell found by Castro-Carrizo el al. (2001a). In addition, an extended intense circumstellar component was found. Although the overall shape of IRC +10420 is approximately spherical, significant departures from that symmetry were detected; an inner bright elongation and important flux deficiencies in the southernmost extended regions are very remarkable.

3. With an LVG code, we modeled the azimuth-averaged brightness distributions obtained for each source. Episodes of strong mass loss have been found to explain the detached shells observed. The deduced mass-loss rates reach much higher values ($\sim 10^{-4}$– $3\,10^{-3}\,M_\odot$ yr$^{-1}$) than usual in AGB stars.

4. Similar episodic mass-loss processes have been found in the two YHGs. Both sources experienced periods of enhanced mass loss and phases with a low or negligible mass-loss rate. The characteristic times of these phases of mass loss are a few thousands of years for both envelopes. Such events may be part of the evolution of YHGs.

5. The YHG nature of AFGL 2343 has sometimes been questioned, mainly because its distance is poorly known (see Sect. 1). The similarities found between the CO emission of IRC +10420 and AFGL 2343, in their extraordinary kinematics, episodic mass-loss phases, and derived time scales, support our assumption about the YHG nature of AFGL 2343. In addition, note that the fast outflows in AFGL 2343 and IRC +10420 can be driven by radiation pressure, contrary to what is found for fast winds in post-AGB stars (B01).

6. It has been proposed that mass loss in YHGs can be due to the presence of an instability region, called *yellow void*, where the normal star pulsation is enhanced by a decrease in the effective gravity force in the photosphere. It is not clear whether this phenomenon could explain the heavy envelopes found around AFGL 2343 and IRC +10420, since the periodicity of their mass loss is not well established and the properties of the circumstellar medium in them are very different from those of the unique YHG that has been proposed so far as bouncing periodically against the yellow void. It is conceivable that the mass ejection of our sources could have been due to a particularly deep, but rare, encounter with the yellow void. But, as we have seen, radiation pressure may also play a relevant role in the mass-loss process, particularly in view of the efficient formation of grains shown by their strong FIR emission. The meaning of the existence of our particularly massive envelopes, within the frame of the general properties of the YHG evolution, therefore remains unknown.

*Acknowledgements.* AC-C acknowledges financial support from the 6th European Community Framework Programme through a Marie-Curie Intra-European Fellowship. The contribution by GQ-L, VB, and JA was supported by the Spanish Ministerio de Ciencia y Tecnología and European FEDER funds, under grants AYA2003-7584 and ESP2003-04957. We acknowledge the IRAM staff for carrying out the observations and particularly P. Hily-Blant and J. Pety for their support in the analysis of the OTF data. We finally thank the anonymous referee for his/her constructive comments.